# A mid-infrared Brillouin laser using ultra-high-Q on-chip resonators

Kiyoung Ko[1, †], Daewon Suk[1, †], Dohyeong Kim[1], Soobong Park[1], Betul Sen[2], Dae-Gon Kim[1], Yingying Wang[3], Shixun Dai[3], Xunsi Wang[3], Rongping Wang[3,4], Byung Jae Chun[5], Kwang-Hoon Ko[5], Peter T. Rakich[2], Duk-Yong Choi[4,*], Hansuek Lee[1,**]

[1]Department of Physics, Korea Advanced Institute of Science and Technology, Daejeon, Republic of Korea

[2]Department of Applied Physics, Yale University, New Haven, CT, USA.

[3]The Research Institute of Advanced Technologies, Ningbo University, Ningbo, China

[4]Laser Physics Centre, Research School of Physics, Australian National University, Canberra, ACT 2601, Australia

[5]Quantum Optics Research division, Korea Atomic Energy Research Institute, Daejeon, Republic of Korea

[†]These authors contributed equally: Kiyoung Ko, Daewon Suk

* duk.choi@anu.edu.au, ** hansuek@kaist.ac.kr

**Abstract**

Ultra-high-Q optical resonators have facilitated recent advancements in on-chip photonics by effectively harnessing nonlinear phenomena providing useful functionalities. While these breakthroughs, primarily focused on the near-infrared region, have extended interest to longer wavelengths holding importance for monitoring and manipulating molecules, the absence of ultra-high-Q resonators in this region remains a significant challenge. Here, we have developed on-chip microresonators with a remarkable Q-factor of 38 million, surpassing previous mid-infrared records by over 30 times. Employing innovative fabrication techniques, including the spontaneous formation of light-guiding geometries during material deposition, resonators with internal multilayer structures have been seamlessly created and passivated with chalcogenide glasses within a single chamber. Major loss factors, especially airborne-chemical absorption, were thoroughly investigated and mitigated by extensive optimization of resonator geometries and fabrication procedures. This allowed us to access the fundamental loss performance offered by doubly purified chalcogenide glass sources, as demonstrated in their fiber form. Exploiting this ultra-high-Q resonator, we successfully demonstrated Brillouin lasing on a chip for the first time in the mid-infrared, with a threshold power of 91.9 μW and a theoretical Schawlow-Townes linewidth of 83.45 Hz, far surpassing carrier phase noise. Our results showcase the effective integration of cavity-enhanced optical nonlinearities into on-chip mid-infrared photonics.

## Introduction

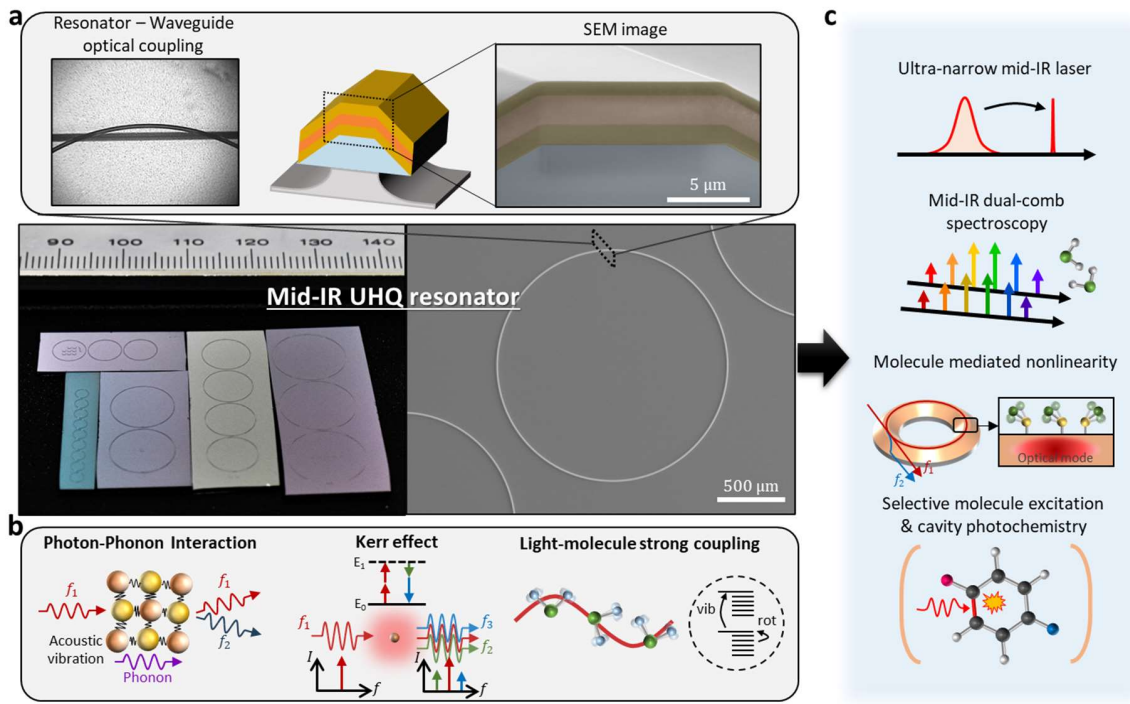

**Fig.1 | Ultra-high-Q on-chip microresonator and its mid-IR application examples. a**, Optical and scanning electron microscope (OM, SEM) images of the mid-IR ultra-high-Q (UHQ) resonator on a chip. The upper-left figure presents the concept of flip-chip coupling between the bus waveguide and the resonator. The SEM image on the right shows a cross-section of the fabricated multilayer UHQ resonator with false coloring for visibility. The lower figures show the OM and SEM images of the actual resonator with varying radii. **b**, Schematic of cavity-enhanced light-matter interactions that can occur using UHQ mid-IR microresonators. **c**, Examples of possible applications in nonlinear optics and molecular physics using the phenomena listed in c.

Over the past decade, ultra-high-Q (UHQ) optical microcavities have become an essential element in on-chip photonics. Significant enhancements in their Q-factors[1–5] have enabled exploration of optical nonlinear processes on a chip, even with the limited pump power of diode lasers. Their geometry can be precisely tailored to control properties such as the free spectral range (FSR) and dispersion, satisfying phase-matching requirements with unprecedented precision beyond free space or fiber optics. The ability to manipulate the coupling geometry provides an additional degree of freedom to control the spectral behaviors of nonlinear dynamics[6]. Leveraging these advancements, phenomena such as the Kerr effect[7], Pockel effect[8], and Brillouin/Raman scatterings[9–11], as well as their ordered combinations[12,13], have been thoroughly investigated and harnessed to implement versatile functionalities within a tiny footprint. These advances in performance encompass ultra-narrow linewidth lasers[3,9], optical wavelength converters[14], and notably, various prominent applications associated with time-frequency precision secured by optical frequency combs[15].

These advancements in on-chip photonics have primarily focused on the near-infrared (near-IR) region, gradually expanding their interest into adjacent longer wavelengths where molecules exhibit fundamental absorption bands. The mid-infrared (mid-IR) region provides a powerful and efficient way to monitor and manipulate the molecule vibrational dynamics, and thus holds significant importance in molecular sensing/spectroscopy[16,17], biochemical imaging[18], and chemical processing[19]. Notably, essential components for these applications, traditionally reliant on large optical setups, have recently been demonstrated on a chip, with examples including tunable light sources through optical parametric oscillation (OPO)[20,21] and soliton pulses directly generated from quantum cascade lasers (QCL)[22,23].

Despite this recent progress, the lack of UHQ resonators in the mid-IR remains a significant bottleneck in on-chip photonics. Multiphonon absorption leads to excessive material loss, particularly hindering conventional material platforms from maintaining their high Q-factor in this region. While materials comprised of heavier elements, such as chalcogenide glasses, alleviate this issue, to date, on-chip waveguides based on them have exhibited higher losses than their fiber optic counterparts with no clear solution in sight. Cavity-oriented nonlinearity adopted by pioneering works in this region, such as mode-locked comb generation[24], can become more efficient and functional with the increase of Q-factors. Furthermore, together with microfabrication-driven precision in controlling the cavity properties, UHQ resonators offer unique opportunities to exploit strong interactions between photons and molecular vibrational modes on a chip, enabling to open new avenues for various research topics such as molecule-induced nonlinear optics[25] and vacuum-field catalysis[19] (Figure 1).

In this paper, we present a new approach to realize UHQ resonators in the mid-IR spectral range, obtaining Q-factors of 38 million which represents a 30-fold enhancement relative to prior works[5]. Through an approach where light-guiding geometries are defined during material deposition, without requiring etch processes, resonators having internal multilayer structures have been seamlessly fabricated and passivated with chalcogenide glasses within a single chamber, thus eliminating interstep contamination. The primary sources of optical loss in the mid-IR are extensively analyzed and mitigated by optimizing the cavity structure and fabrication conditions, including material absorption, surface scattering, and surface-adsorbed airborne chemical absorption whose importance is revealed here. Leveraging the high Q-factor and controllability of FSR afforded by microfabrication, stimulated Brillouin lasing is demonstrated for the first time in the mid-IR, exhibiting a threshold power of 91.9 µW, well below the output power of on-chip OPO lasers and QCLs, and a theoretical Schawlow-Townes linewidth of 83.45 Hz, significantly surpassing that of commercialized QCLs.

**Results**

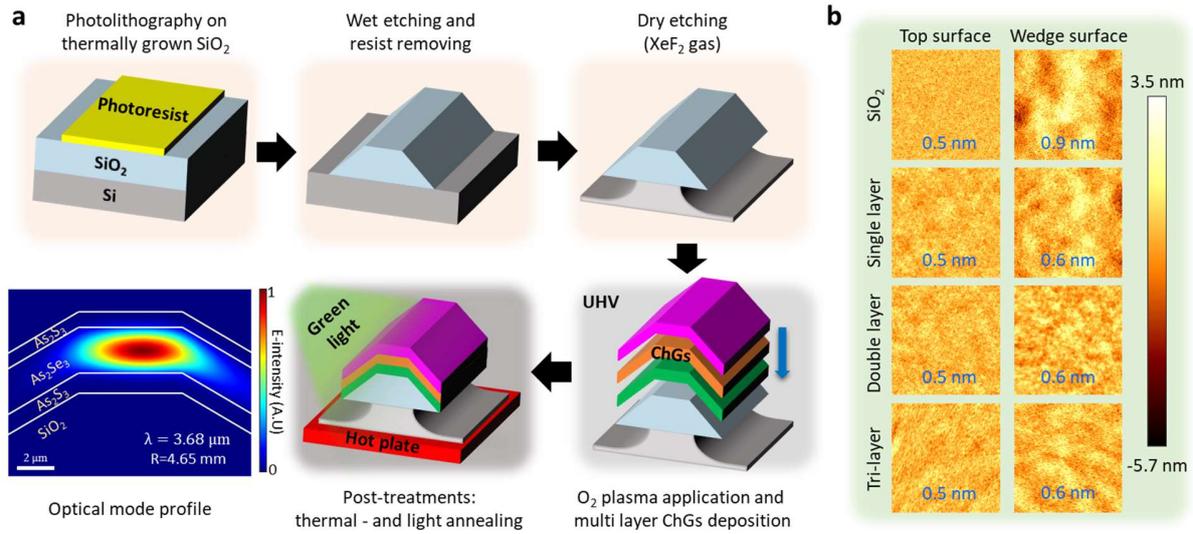

**Fig.2 | Fabrication of a multi-layer chalcogenide glass resonator. a,** The fabrication steps for the proposed mid-IR UHQ resonator are depicted, with a detailed explanation given in the Methods. The different colored ChG layers represent the bottom cladding - (green), the upper cladding - (pink), and the core layer (orange). The bottom left corner shows an example of the calculated optical mode in a resonator having the same geometric parameters as seen in Figure 3b. **b,** Surface roughness measured by AFM at the top and wedge surfaces of the trapezoid. Roughness data are shown for the $SiO_2$ substructure and after sequentially depositing one to three layers of ChG on that structure. The numbers show the root mean square (rms) roughness measured over a 3×3 µm² area. The thickness and material of each layer are as follows: first layer: 1.1 µm $As_2S_3$, second layer: 2.2 µm $As_2Se_3$, third layer: 0.9 µm $As_2S_3$.

**Design and fabrication of on-chip chalcogenide glass resonators in the mid-IR.** As a material platform for UHQ resonators, we chose chalcogenide glass (ChG) because of its wide transmission window covering the entire mid-IR spectrum, large nonlinearities, and tailorable optical properties provided by the material composition[26,27]. Previous on-chip ChG microresonators exhibited Q-factors as high as 6 x 10⁵ [28] in the mid-IR, which is significantly lower than the Q-factor of 1.9 x 10⁹ converted from the lowest loss (12 dB/km[29]) exhibited by chalcogenide based fiber optics. This substantial difference has been attributed to practical factors that have yet to be fully optimized on a chip such as surface roughness, the presence of impurities in the material, and chemicals adsorbed on the surface. This study thoroughly investigates and addresses these factors.

To push the performance of such waveguides closer to the intrinsic losses supported by ChG materials, optical losses induced by surface scattering have been reduced to a fundamental level, limited by the molecular-scale roughness of the ChG film surface. Conventional methods for shaping light-guiding structures involve subtractive techniques such as etching[30,31] and lift-off[32,33] to create spatial optical thickness variation, often resulting in rough side walls that cause high scattering-induced loss. Furthermore, owing to the amorphous nature and stoichiometric irregularities of ChG, attaining smooth side walls has been even more challenging[34]. This problem was completely resolved by employing a method where the light-guiding geometry is spontaneously formed during material deposition, eliminating the need for subsequent subtractive processes[35]. The optical thickness distribution is created seamlessly, without side walls, along the platform by directionally depositing ChG on a wet-etched $SiO_2$ substructure having a trapezoidal cross-section, as depicted in Figure 2a with more details in the Methods. The optical modes are confined to the top flat region of the deposited film, which is optically thicker than the sloped region.

To adapt this near-IR-proven technique for mid-IR applications, a bottom cladding layer made of ChG is incorporated to counter the Q-factor reduction caused by the excessive material loss of $SiO_2$ in the mid-IR (460 dB/m at 3.7 µm)[36]. As shown in Figure 2a, the optical modes are contained within the core layer, effectively separated from the $SiO_2$ platform by the bottom cladding layer. The specific

ChGs chosen for the core and cladding were $As_2Se_3$ ($n_{core}$: 2.77 at 3.7 μm) and $As_2S_3$ ($n_{clad}$: 2.41 at 3.7 μm), respectively. The proposed resonator featuring an internal multilayer structure can be easily fabricated by depositing ChGs successively in a single high vacuum (HV, < 5×10$^{-8}$ torr) thermal evaporation chamber. In addition, the deposition reduces the surface roughness that is initially present on the bottom $SiO_2$ platform to sub-nanometer levels at the core-cladding interfaces, as confirmed by atomic force microscopy (AFM) measurements shown in Figure 2b.

To further approach the loss limits of ChG materials, absorption losses resulting from in-process contamination and impurities originating from the source materials, which have been extensively studied in low-loss mid-IR ChG fibers[29,37,38], have been minimized. Chemical contamination during fabrication was inherently blocked by sequentially depositing the internal multilayers in a single HV chamber, in contrast to traditional microfabrication methods that involve multiple steps of deposition and etching. In addition, prior to deposition, the $SiO_2$ substructure is cleansed with $O_2$ plasma inside the deposition chamber to remove surface contaminants. Impurities originating from the sources in the deposited films were suppressed by selecting highly purified ChGs, similar to those used for mid-IR fibers, as the deposition sources. The sources are purified to systematically eliminate hydrogen and oxygen-related impurities, as detailed in the Methods. In particular, a double-purified source[37] was used for core deposition, which led to a notable improvement in the Q-factor, as discussed in Supplement 1C.

In addition, a mid-IR passivation material that can preserve the ultra-high-Q while providing sufficient degradation resistance was applied and scrutinized. Although fluoropolymer films and atomic layer deposited (ALD) films have traditionally been utilized for passivating mid-IR devices, they introduce significant losses owing to their intrinsic absorption band[39] or poor chemical purity[40]. Here, we verified Ge-As-Se, a ChG variant known for its high mid-IR transmittance[38] and stability under ambient conditions, as a passivation layer for mid-IR UHQ resonators. Examination revealed that a 10 nm thick Ge-As-Se film, deposited consecutively within the same HV chamber, effectively prevented crystallization of the ChG core and claddings (Supplement 1B), without causing an observable change in loss (Supplement 3). As a final step, a post-treatment by thermal and light annealing under a medium vacuum condition (<1 mTorr) is applied to minimize local irregularities inside the deposited films, resulting in an additional increase in the Q-factor, as presented in the Methods and Supplement 1C.

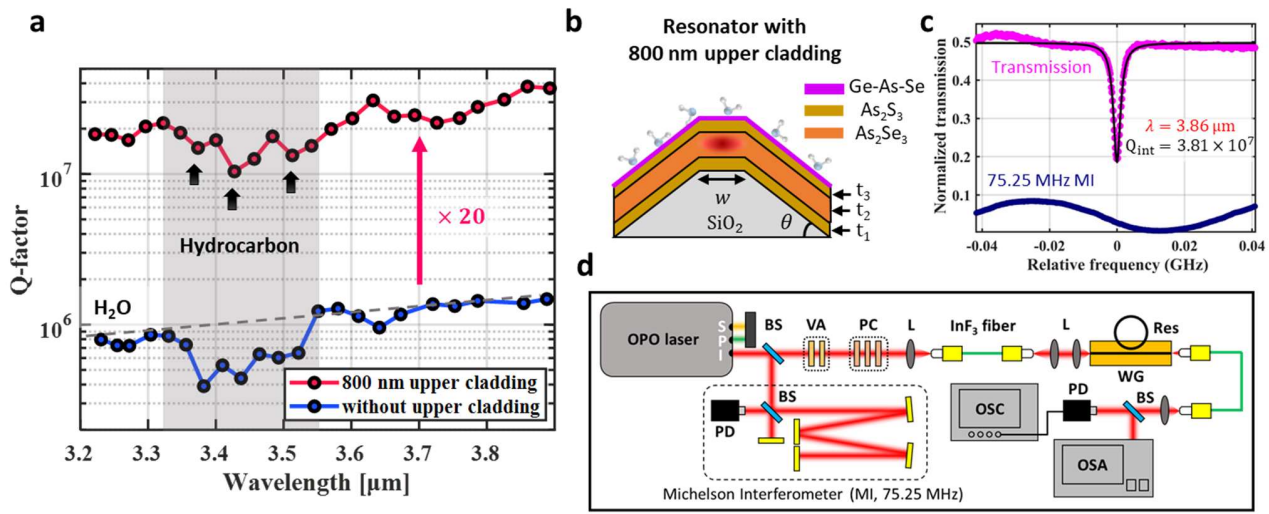

**Fig. 3 | Q-factor spectra and measurement set-up. a,** Q-factor spectrum of two different resonators, with and without an upper cladding, measured at wavelengths of $3.2 - 3.9$ μm. The grey scaled region at the $3.33 - 3.55$ μm wavelength indicates the hydrocarbon (C-H) absorption band, with their characteristic peaks indicated with black arrows. The grey dashed line is given as a guideline for the effect of the $H_2O$ absorption tail. **b,** Cross-section drawing of the resonator with an upper cladding. The geometric parameters are as follows: $w = 6.07$ μm, $\theta = 28°$, $t_1 = 1.05$ μm, $t_2 = 2.4$ μm, $t_3 = 0.8$ μm. **c,** Lorentzian fit of the record Q-factor measured at 3.86 μm wavelength on the same resonator. The magenta line indicates the resonator transmission spectrum, while the navy line is the interferogram of the calibrated Michelson interferometer (MI, 75.25 MHz), depicted in Figure d, used as a frequency reference. **d,** A drawing of the mid-IR cavity coupling setup. A continuously tunable optical parametric oscillator (OPO) laser was used as the pump source. Details about the setup and interferometer are given in the Methods. BS: beam splitter, VA: variable attenuator, PC: polarization controller, L: lens, Res: resonator, WG: bus waveguide, PD: photodetector, OSC: oscilloscope, OSA: optical spectrum analyzer.

**Post-fabrication loss from surface-adsorbed molecules.** On-chip resonators, which are typically unclad or clad with considerably thinner layers than optical fibers, are susceptible to absorption caused by molecules that continuously adhere to the surface from the surrounding air. This post-fabrication degradation, often disregarded in the near-IR, is exacerbated in the mid-IR because of the strong absorption associated with fundamental molecular vibrations[27,39]. Therefore, besides the aforementioned refinements in device design and fabrication, it is crucial to quantitatively assess the absorption loss resulting from surface-adsorbed airborne chemicals and suppress it.

This loss factor was thoroughly analyzed by measuring the Q-factor of resonators with various upper cladding thicknesses across a wide spectral range. Figure 3a depicts the measured Q-factor, covering wavelengths from 3.2 μm to 3.9 μm, for resonators with two contrasting cladding thicknesses, 0 (no cladding) and 800 nm, respectively, as illustrated in Figure 3b. It is obvious that a cladding thickness of 800 nm is enough to enhance the overall Q-factor spectrum by approximately twenty times. The relatively lower Q-factors of the resonator without cladding are not constrained by the fabrication issues described in the previous section but rather are attributed to molecular absorption, as evidenced by the Q-factor exceeding $10^7$ at 1550 nm, which is discussed in Supplement 2. Airborne impurities adhere to the outermost surfaces of both resonators in similar concentrations because of identical surface properties defined by the Ge-As-Se passivation layer, which remains unaffected by the upper cladding thickness. Consequently, within the regime where surface molecular absorption dominates, the loss (Q-factor) is linearly (inversely) proportional to the outermost surface electric field intensity, diminishing exponentially with increasing cladding thickness. This trend was validated experimentally using resonators with various upper cladding thicknesses (Supplement 3), enabling quantification of surface molecular absorption. By incorporating this loss factor into the previously established loss calculation based on material absorption and surface scattering, the Q-factors of a given resonator geometry could be accurately predicted over wavelengths from 3 to 4 μm, as described in the Methods. This analysis, delineating the contribution of each loss factor, offers insight for further enhancing Q-factors (Supplement 5).

Precise measurement of Q-factor spectra, calibrated against a frequency reference (see Figure 3c, d and Methods), provides sufficient resolution to discern the distinct contributions of specific molecular groups to absorption loss, based on their unique spectral fingerprints. Within the measurement range shown in Figure 3a, two primary contaminants are identified: water and hydrocarbons (C – H). The extended tail of the well-known broad $H_2O$ absorption band, centered at approximately 2.9 μm[27,39], imposes a limitation on the overall Q-factor, as indicated by the grey dotted line in Figure 3a. Hydrocarbons, on the other hand, induce sharp decreases in Q-factors, corresponding to their characteristic absorption peaks at 3.38, 3.42, and 3.52 μm[39]. This highlights the importance of carefully managing airborne chemicals that exhibit significant absorption within the spectral band aimed for UHQ resonator operation, such as regulating surface adhesion and cladding thickness.

It is worth highlighting that the intrinsic Q-factor ($Q_{int}$) of $3.81 \times 10^7$ at a wavelength of 3.86 μm (Figure 3c) represents a remarkable breakthrough in the mid-IR range, surpassing all previously reported values by a factor exceeding 30, including those obtained using ChG[28] or any other on-chip materials such as silicon[5], silicon nitride[41], germanium[42], and III−V semiconductors[43]. This achievement significantly reduces the required pump power for most nonlinearities by approximately a thousand times ($\propto 1/Q^2$). This in turn enables the realization of numerous optical functionalities on a chip without being constrained by the absence of commercial amplifiers in this region. Furthermore, the propagation loss of 0.52 dB/m converted from this Q-factor is only twice as much as that of the state-of-the-art ChG fiber made from the same ingots, 0.25 dB/m[37], thus verifying that all analyzed loss factors have been sufficiently minimized to achieve performance comparable to the fiber. Figure 4 shows a comprehensive visual comparison, including the mentioned loss values.

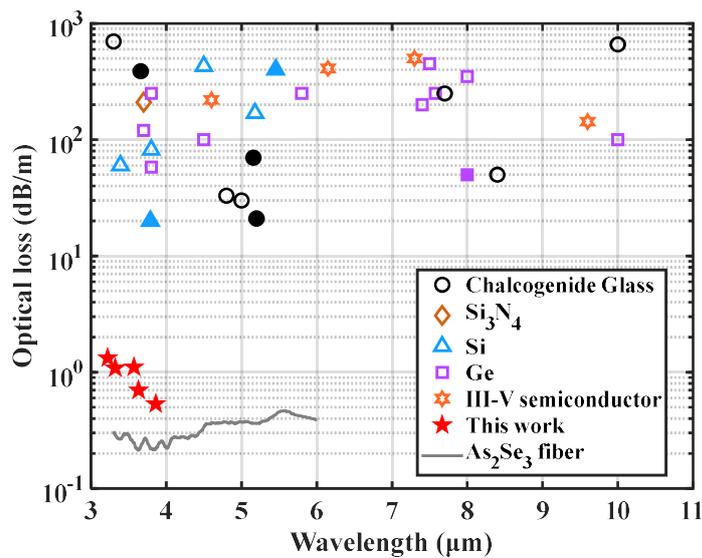

**Fig. 4 | Comparison of optical losses with different material on-chip platforms.** The propagation losses of diverse mid-IR on-chip resonators and waveguides are compared with this work. Different markers indicate different material platforms: chalcogenide glasses (black, circle), silicon nitride (brown, diamond), silicon (blue, triangle), germanium (purple, square), and III–V semiconductors (orange, hexagram). The solid symbols represent resonators, and the open symbols indicate waveguides. The optical losses of this work are converted from the record Q-factors shown in Figure 3a. The loss of state-of-the-art $As_2Se_3$ fiber fabricated using identical source materials in this work is drawn as a grey line. All references are given in Supplement 8.

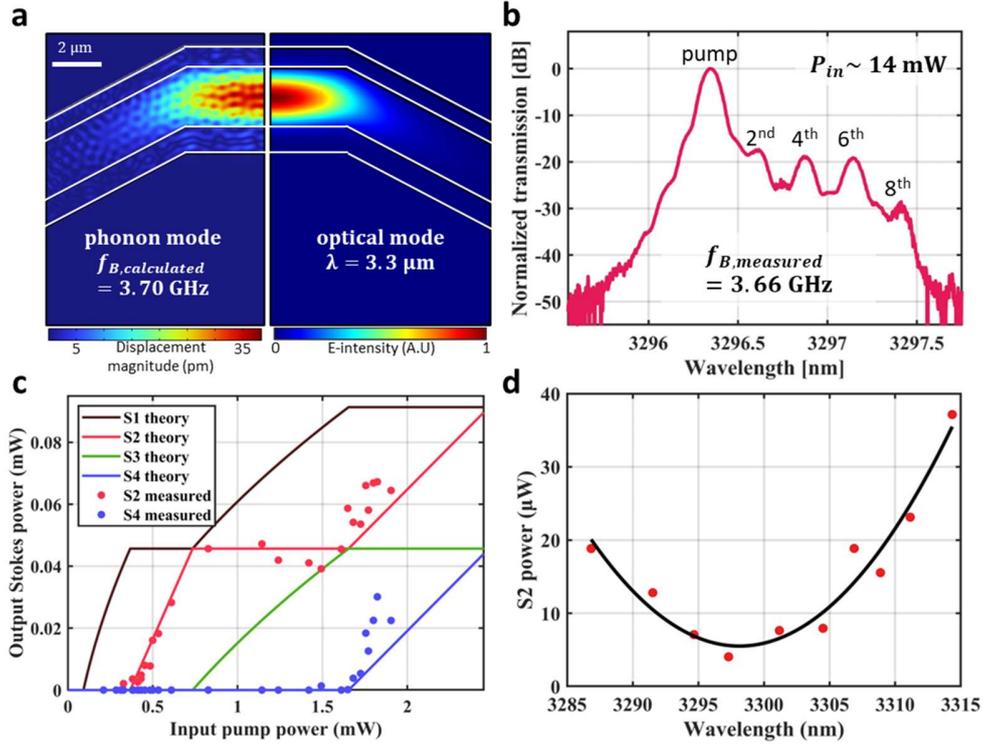

**Fig. 5 | Mid-IR stimulated Brillouin lasing (SBL). a,** Intensity profile of the acoustic and optical eigenmodes responsible for the stimulated Brillouin scattering (SBS) signal generated at 3.3 μm wavelength. Only the left and right halves of each mode are shown, as both modes are bilaterally symmetric. **b,** Optical power spectrum showing SBL operation up to the 8$^{th}$ order, measured at the transmission port using an optical spectrum analyzer. **c,** Measured and calculated power dynamics curve of the 2$^{nd}$ and 4$^{th}$ Stokes wave (S2, S4). The x and y axes of the calculated curves (solid lines) were both multiplied by a scaling factor of β=1.43 to fit the measured values. The 1$^{st}$ Stokes wave threshold power of 91.9 μW was extracted from the fit. **d,** Clamped power of the S2 plotted as a function of the pump wavelength, measured to calculate the Brillouin gain bandwidth.

**An on-chip Brillouin laser in the mid-IR.** Taking advantage of the ultra-high Q-factor of the microresonators, a Brillouin laser was developed for the first time in the mid-IR. The employed resonator has a cross-sectional design identical to that shown in Figure 3b. As confirmed in Figure 5a, this particular design confines, within the As$_2$Se$_3$ core layer, the fundamental optical mode and the phonon mode that yields the highest Brillouin gain. These modes were identified using the finite element method, treating the device as a straight waveguide, resulting in a modal overlap factor of around 0.96[44]. A full-vectorial model of the Brillouin coupling[45] predicts an appreciable Brillouin gain, $G_B$, of 59.90 m$^{-1}$ W$^{-1}$. This is attributed to the high overlap factor and the inherently large Brillouin gain coefficient of As$_2$Se$_3$[46] exceeding that of conventional on-chip optical materials such as SiO$_2$[47], Si$_3$N$_4$[48], and Si[47] by over 100 times. All parameters for the Brillouin gain analysis are provided in Supplement 7. Throughout this analysis and subsequent experiments, the pump wavelength was set at 3.3 μm, considering the measurement capability of the available optical spectrum analyzer (OSA, Yokogawa AQ6376, with a wavelength range of 1.5 − 3.4 μm), despite the observed increase in the Q-factors at longer wavelengths, as depicted in Figure 3a. The radius of the resonator was set to 4.65 mm to match its free spectral range (FSR) with the calculated Brillouin frequency shift, $f_B$, of 3.70 GHz.

The generation of a Brillouin lasing signal in the microcavity was experimentally verified by observing a cascading process extending to the 8$^{th}$ Stokes wave, as shown in Figure 5b. This result was achieved with a waveguide-coupled input pump power of around 14 mW. Throughout the measurement, the forward-traveling signal, composed of even-order Stokes waves, was captured by an OSA at the transmission port of the coupled waveguide since optical circulators were unavailable within this wavelength range. The $Q_{int}$ of the pumped mode was $2.19 \times 10^7$. The $f_B$, extracted from the optical power spectrum, was 3.66 GHz, which corresponds closely to the calculated value within the OSA resolution limit of 0.003 nm, or equivalently, 82.8 MHz.

The lasing process exhibited a remarkably low threshold, as determined by analyzing the power dynamics involved. In Figure 5c, the measured power of the second (S2) and fourth (S4) Stokes waves is represented by red and blue dots, respectively, plotted against the input pump power. Power dynamic curves analytically derived by cascade-order Brillouin laser theory[49], represented by solid lines in the figure, show an excellent match with the measured values. For the best fitting, a scaling factor (β) of 1.43 was multiplied by both the input and output power of the theoretical curves, compensating for the measurement error in waveguide-to-fiber coupling loss, as explained in the Methods. The lasing thresholds, obvious in Figure 5c, were identified as 0.367 mW for S2 and 1.65 mW for S4.

Importantly, the threshold for S1 was calculated to be 91.9 μW from this analysis. This value is comparable to the lowest threshold observed in on-chip Brillouin lasers in the near-IR, which was 40 μW, achieved using a SiO$_2$ resonator with a $Q_{int}$ of $3 \times 10^8$ [10].

To further characterize the laser properties, the Brillouin phonon gain bandwidth was measured using the pump wavelength detuning method[10]. By detuning the pump wavelength from the Brillouin gain peak wavelength, a discrepancy between $f_B$ and cavity FSR increases, which decreases the gain in accordance with its spectral profile. In Figure 5d, the clamped power of the 2$^{nd}$ Stokes wave, which is inversely proportional to the gain coefficient, is plotted against the pump wavelength. The full-width-half-maximum (FWHM) of its quadratic fit, depicted by the black line, is 13.91 nm. The phonon gain bandwidth of 14.51 MHz, or equivalently the phonon Q-factor of 248.5, was obtained by multiplying this FHWM by the frequency mismatch per pump wavelength detuning (Hz/nm), calculated considering the frequency pulling effect[10,50]. Details of this calculation are provided in Supplement 6.

The improvement in spectral purity achieved through the Brillouin lasing process was estimated theoretically using established models for Schawlow–Townes (ST) narrowing in Brillouin lasers. Since the measured phonon bandwidth ($\Gamma = 2\pi \times 14.51$ MHz) is comparable to the photon decay rate ($\gamma = \sim 2\pi \times 7$ MHz) of the cavity, the ST equation without adiabatic approximation ($\gamma/\Gamma \ll 1$) was employed to calculate the linewidth[50], $\Delta\nu_0 = \left(\frac{\Gamma}{\gamma+\Gamma}\right)^2 \frac{\hbar\omega_L^3 n_{th}}{4\pi Q_T Q_{ex} P_{SBL}}$. Using the measured parameters for optical and acoustic damping, as described in the Methods, yielded an ST linewidth for the clamped Stokes wave of 83.45 Hz. Remarkably, this value is more than a thousand times narrower than the linewidth of the pump laser source (Argos SF-15 module C, linewidth < 1 MHz) or commercialized quantum cascade lasers (Alpes Lasers, ~1 MHz). In conjunction with quantum cascade lasers, this advance offers significant potential to develop ultra-narrow linewidth lasers on a chip encompassing a broad range of the mid-infrared spectrum.

**Discussion**

In conclusion, we achieved ultra-high Q-factors in the mid-IR on-chip resonators by developing designs and fabrication methods optimized to individually address all prominent loss factors in mid-IR integrated photonics. The proposed design, featuring internal multi-layers that could be fabricated without a direct etch process, effectively mitigates material absorption and scattering losses that traditionally hindered mid-IR Q-factors. By doing so, post-fabrication loss due to surface-adsorbed airborne molecules, which was previously indiscernible, was unveiled, quantitatively analyzed, and then addressed by tuning the geometry for the first time. The combined efforts resulted in a record Q-factor of 38 million, equivalent to a propagation loss of 0.52 dB/m, significantly outperforming other on-chip mid-IR devices and approaching the losses of state-of-the-art ChG fiber (Figure 4). This ultra-high Q-factor led to the first realization of a mid-infrared Brillouin laser with low threshold power (91.9 μW) and an estimated Schawlow-Townes linewidth (83.45 Hz), demonstrating the potential for highly efficient on-chip ultra-narrow linewidth lasers in the mid-IR. The loss analysis presented here offers insights for reducing the optical loss of other mid-IR material platforms, suggesting possibilities for low-loss mid-IR optical circuits like those in the near-IR. Furthermore, ultra-high-Q resonators, leveraging their fabrication precision and geometric tunability, offer the potential for cavity-enhanced light-molecule interaction on a chip, presenting unique research opportunities in nonlinear optics and molecular science in the mid-IR.

**References**

1. Wu, L. *et al.* Greater than one billion Q factor for on-chip microresonators. *Opt. Lett., OL* **45**, 5129–5131 (2020).
2. Liu, J. *et al.* High-yield, wafer-scale fabrication of ultralow-loss, dispersion-engineered silicon nitride photonic circuits. *Nat Commun* **12**, 2236 (2021).
3. Jin, W. *et al.* Hertz-linewidth semiconductor lasers using CMOS-ready ultra-high-Q microresonators. *Nat. Photonics* **15**, 346–353 (2021).
4. Zhang, M., Wang, C., Cheng, R., Shams-Ansari, A. & Lončar, M. Monolithic ultra-high-Q lithium niobate microring resonator. *Optica, OPTICA* **4**, 1536–1537 (2017).
5. Miller, S. A. *et al.* Low-loss silicon platform for broadband mid-infrared photonics. *Optica* **4**, 707 (2017).
6. Gong, Z. *et al.* Photonic Dissipation Control for Kerr Soliton Generation in Strongly Raman-Active Media. *Phys. Rev. Lett.* **125**, 183901 (2020).
7. Kippenberg, T. J., Gaeta, A. L., Lipson, M. & Gorodetsky, M. L. Dissipative Kerr solitons in optical microresonators. *Science* **361**, eaan8083 (2018).
8. Wang, C. *et al.* Integrated lithium niobate electro-optic modulators operating at CMOS-compatible voltages. *Nature* **562**, 101–104 (2018).
9. Gundavarapu, S. *et al.* Sub-hertz fundamental linewidth photonic integrated Brillouin laser. *Nature Photon* **13**, 60–67 (2019).
10. Li, J., Lee, H., Chen, T. & Vahala, K. J. Characterization of a high coherence, Brillouin microcavity laser on silicon. *Opt. Express, OE* **20**, 20170–20180 (2012).
11. Spillane, S. M., Kippenberg, T. J. & Vahala, K. J. Ultralow-threshold Raman laser using a spherical dielectric microcavity. *Nature* **415**, 621–623 (2002).
12. Do, I. H. *et al.* Self-stabilized soliton generation in a microresonator through mode-pulled Brillouin lasing. *Opt. Lett., OL* **46**, 1772–1775 (2021).
13. Do, I. H. *et al.* Spontaneous soliton mode-locking of a microcomb assisted by Raman scattering. *Opt. Express, OE* **31**, 29321–29330 (2023).
14. Stone, J. R. *et al.* Wavelength-accurate nonlinear conversion through wavenumber selectivity in photonic crystal resonators. *Nat. Photon.* **18**, 192–199 (2024).
15. Diddams, S. A., Vahala, K. & Udem, T. Optical frequency combs: Coherently uniting the electromagnetic spectrum. *Science* **369**, eaay3676 (2020).
16. Spaun, B. *et al.* Continuous probing of cold complex molecules with infrared frequency comb spectroscopy. *Nature* **533**, 517–520 (2016).
17. Abbas, M. A. *et al.* Time-resolved mid-infrared dual-comb spectroscopy. *Sci Rep* **9**, 17247 (2019).
18. Zhang, D. *et al.* Depth-resolved mid-infrared photothermal imaging of living cells and organisms with submicrometer spatial resolution. *Science Advances* **2**, e1600521 (2016).
19. Garcia-Vidal, F. J., Ciuti, C. & Ebbesen, T. W. Manipulating matter by strong coupling to vacuum fields. *Science* **373**, eabd0336 (2021).
20. Roy, A. *et al.* Visible-to-mid-IR tunable frequency comb in nanophotonics. *Nat Commun* **14**, 6549 (2023).
21. Hwang, A. Y. *et al.* Mid-infrared spectroscopy with a broadly tunable thin-film lithium niobate optical parametric oscillator. *Optica, OPTICA* **10**, 1535–1542 (2023).

22. Meng, B. et al. Dissipative Kerr solitons in semiconductor ring lasers. *Nat. Photon.* **16**, 142–147 (2022).
23. Opačak, N. et al. Nozaki–Bekki solitons in semiconductor lasers. *Nature* **625**, 685–690 (2024).
24. Yu, M., Okawachi, Y., Griffith, A. G., Lipson, M. & Gaeta, A. L. Mode-locked mid-infrared frequency combs in a silicon microresonator. *Optica* **3**, 854 (2016).
25. Shen, X., Choi, H., Chen, D., Zhao, W. & Armani, A. M. Raman laser from an optical resonator with a grafted single-molecule monolayer. *Nat. Photonics* **14**, 95–101 (2020).
26. Eggleton, B. J., Luther-Davies, B. & Richardson, K. Chalcogenide photonics. *Nature Photon* **5**, 141–148 (2011).
27. Adam, J.-L. & Zhang, X. *Chalcogenide Glasses: Preparation, Properties and Applications*. (Woodhead publishing, 2014).
28. Lin, H. et al. High-Q mid-infrared chalcogenide glass resonators for chemical sensing. in *2014 IEEE Photonics Society Summer Topical Meeting Series* 61–62 (IEEE, 2014).
29. Churbanov, M. F., Skripachev, I. V., Snopatin, G. E., Ketkova, L. A. & Plotnichenko, V. G. The problems of optical loss reduction in arsenic sulfide glass IR fibers. *Optical Materials* **102**, 109812 (2020).
30. Zhang, B., Xia, D., Zhao, X., Wan, L. & Li, Z. Hybrid-integrated chalcogenide photonics. *Light: Advanced Manufacturing* **4**, 503–518 (2024).
31. Du, Q. et al. Low-loss photonic device in Ge–Sb–S chalcogenide glass. *Opt. Lett., OL* **41**, 3090–3093 (2016).
32. Hu, J. et al. Si-CMOS-compatible lift-off fabrication of low-loss planar chalcogenide waveguides. *Opt. Express, OE* **15**, 11798–11807 (2007).
33. Hu, J. et al. Optical loss reduction in high-index-contrast chalcogenide glass waveguides via thermal reflow. *Opt. Express, OE* **18**, 1469–1478 (2010).
34. Choi, D.-Y., Madden, S., Rode, A., Wang, R. & Luther-Davies, B. Nanoscale phase separation in ultrafast pulsed laser deposited arsenic trisulfide (As2S3) films and its effect on plasma etching. *Journal of Applied Physics* **102**, 083532 (2007).
35. Kim, D.-G. et al. Universal light-guiding geometry for on-chip resonators having extremely high Q-factor. *Nature communications* **11**, 5933 (2020).
36. Kitamura, R., Pilon, L. & Jonasz, M. Optical constants of silica glass from extreme ultraviolet to far infrared at near room temperature. *Appl. Opt., AO* **46**, 8118–8133 (2007).
37. Wang, J. et al. Se-H-free $As_2Se_3$ fiber and its spectral applications in the mid-infrared. *Opt. Express, OE* **30**, 24072–24083 (2022).
38. Tang, Z. et al. Low loss Ge-As-Se chalcogenide glass fiber, fabricated using extruded preform, for mid-infrared photonics. *Opt. Mater. Express* **5**, 1722 (2015).
39. Ma, P. et al. Low-loss chalcogenide waveguides for chemical sensing in the mid-infrared. *Opt. Express* **21**, 29927 (2013).
40. Verlaan, V., van den Elzen, L. R. J. G., Dingemans, G., van de Sanden, M. C. M. & Kessels, W. M. M. Composition and bonding structure of plasma-assisted ALD Al2O3 films. *physica status solidi c* **7**, 976–979 (2010).
41. Tai Lin, P., Singh, V., Kimerling, L. & Murthy Agarwal, A. Planar silicon nitride mid-infrared devices. *Applied Physics Letters* **102**, 251121 (2013).
42. Ren, D., Dong, C., Addamane, S. J. & Burghoff, D. High-quality microresonators in the longwave infrared based on native germanium. *Nat Commun* **13**, 5727 (2022).
43. Montoya, J. et al. Integration of quantum cascade lasers and passive waveguides. *Applied Physics Letters* **107**, 031110 (2015).
44. Suh, M.-G., Yang, Q.-F. & Vahala, K. J. Phonon-Limited-Linewidth of Brillouin Lasers at Cryogenic Temperatures. *Phys. Rev. Lett.* **119**, 143901 (2017).
45. Qiu, W. et al. Stimulated Brillouin scattering in nanoscale silicon step-index waveguides: a general framework of selection rules and calculating SBS gain. *Opt. Express, OE* **21**, 31402–31419 (2013).
46. Abedin, K. S. Observation of strong stimulated Brillouin scattering in single-mode $As_2Se_3$ chalcogenide fiber. *Opt. Express, OE* **13**, 10266–10271 (2005).
47. Smith, M. J. A. et al. Metamaterial control of stimulated Brillouin scattering. *Opt. Lett., OL* **41**, 2338–2341 (2016).
48. Gyger, F. et al. Observation of Stimulated Brillouin Scattering in Silicon Nitride Integrated Waveguides. *Phys. Rev. Lett.* **124**, 013902 (2020).
49. Behunin, R. O., Otterstrom, N. T., Rakich, P. T., Gundavarapu, S. & Blumenthal, D. J. Fundamental noise dynamics in cascaded-order Brillouin lasers. *Physical Review A* **98**, 023832 (2018).
50. Yuan, Z., Wang, H., Wu, L., Gao, M. & Vahala, K. Linewidth enhancement factor in a microcavity Brillouin laser. *Optica, OPTICA* **7**, 1150–1153 (2020).

## Methods

**Device fabrication process.** The multilayer ChG resonators were fabricated based on an ultra-high quality $SiO_2$ trapezoidal substructure[35,51] following the sequence shown in Figure 2a. The $SiO_2$ structure was fabricated from an 8 μm thick $SiO_2$ layer thermally grown in an oxidation furnace (Tytan Mini 1800, TYSTAR). Photolithography and wet etching were applied to form the trapezoidal geometry, with its wedge angle controlled by a hexamethyldisilazane (HMDS) vapor treatment. After removing the photoresist, the Si substrate was isotropically etched using $XeF_2$ gas ($XeF_2$ etching system, Teraleader Co. Ltd.) to isolate the optical mode from it. The devices were then loaded into a HV thermal evaporator (<5×$10^{-8}$ torr) (KVE-T2000, Korea Vacuum Tech) and cleaned with $O_2$ plasma. Multiple layers of ChG were then sequentially deposited, including a 10 nm thick Ge-As-Se passivation layer. As a final step, post-treatments, thermal annealing below the glass transition temperature ($T_g$) and light annealing using a 540 nm wavelength lamp, were applied in a medium vacuum (<1 mTorr) for 12 hours. The similar $T_g$ of $As_2S_3$ and $As_2Se_3$ is beneficial for the thermal annealing process. This procedure realigns the internal bonding structures of the films, bringing them closer to their bulk state[52] and thereby reducing optical loss. More details of the fabrication process are presented in Supplements 1A, B, and C.

**Minimizing film incorporates impurities.** To minimize source-borne impurities in the deposited films, high-purity ChG ingots were used as evaporation sources. Commercially available materials from Amorphous Materials Inc. were used for the cladding ($As_2S_3$, AMTIR-6) and passivation ($Ge_{33}As_{12}Se_{55}$, AMTIR-1) layers. For the core layer, which has the greatest effect on the resonator impurity loss, a specially purified homemade $As_2Se_3$ ingot (Ningbo University) used for extruding ultra-low-loss optical fibers[37] was chosen. The simplified purification process for the $As_2Se_3$ ingot is as follows. A single purification cycle is composed of three steps. First, Se-H impurities are removed by dynamic distillation using a $TeCl_4$ hydrogen removal agent, which extracts hydrogen impurities in the form of HCl gas. After air-quenching and annealing, a static distillation using Mg is applied to remove -O and -OH-related impurities. Finally, dynamic distillation with $TeCl_4$ is applied once more to eliminate hydrogen impurities reintroduced from the oxygen-removed-OH impurities. The $As_2Se_3$ sources used for our highest-quality resonators were further purified by performing the whole cycle twice. Additionally, all evaporation sources were sonicated to remove surface contaminants and then dehydrated before being loaded into the chamber. More details are introduced in Supplements 1C and D.

**Mid-IR Q-factor measurement method.** In this study, the Q-factor is attained from the time trace of resonance measured by the laser frequency scanning technique, calibrated with the characteristic frequency of a free-space interferometer acting as a reference. This method requires the use of a continuously tunable laser with a much narrower linewidth compared to the target resonator. The Lockheed Martin Aculight Argos SF-15 laser (module C) used in all our experiments guarantees a linewidth below 1 MHz, which is narrow enough to measure the linewidth of cavities with $Q_{int} < 10^8$.

The measured time trace was accurately converted to a frequency scale using a home-built and calibrated free-space Michelson interferometer (MI). The optical path difference of the two interferometer arms was 3.96 m, resulting in a ~75 MHz interference pattern. The MI was calibrated by comparing its interferogram with a reference fiber Mach-Zehnder interferometer (MZI) at 1.55 μm wavelength. Both interferometers were connected to a single 1.55 μm laser, and the number of interference fringes was simultaneously measured during a wavelength scan. By scaling the reference MZI frequency with the number of fringes, the frequency of MI at 1.55 μm wavelength was calculated to be 75.25 MHz. The same frequency value was used for calculating Q-factors at mid-IR wavelengths, ignoring the chromatic dispersion of air and free-space optics used for the MI. The temperature and humidity of the free-space MI setup, which affect the refractive index of air, were kept nearly constant during the entire experiment using a thermo-hygrostat (T = 18 °C, H = 15 %).

Optical power propagating in the bus waveguide was coupled to the trapezoid resonator using a flip-chip coupling scheme[35]. The bus waveguides were fabricated as bilayers without an upper cladding to attain maximum coupling strength with the resonator. The index difference between the waveguide and resonator was compensated using a two-point coupling scheme[53], which allowed us to achieve nearly critical coupling on all measured wavelengths.

**Predicting Q-factor spectral tendency of resonators with varying geometries.** The theoretical Q-factors of resonators with varying structural parameters were predicted accounting for all the previously discussed loss factors, including bottom $SiO_2$ absorption, surface scattering, and surface chemical absorption. $SiO_2$ absorption was calculated by optical mode simulations, utilizing material absorption coefficients sourced from a reference[36]. Scattering losses were determined using the volume current method[54,55], employing surface roughness measured by AFM in Figure 2b. Surface chemical absorption by $H_2O$ was estimated by extracting the absorption coefficient from the Q-factor of a reference resonator and subsequently multiplying it by the simulated surface E-field intensity, as outlined in Supplement 4. Compared with the measured Q-factors in Figure 3a, the calculated values correspond well, as illustrated in Figure 6. The deviation of measured Q-factors from the calculation stems from loss factors that were unaccounted for in this simulation, specifically hydrocarbon absorption within the 3.35 − 3.6 μm wavelength range, and possibly S-H impurity absorption within the 3.65 − 3.85 μm[27] range, originating from the $As_2S_3$ cladding.

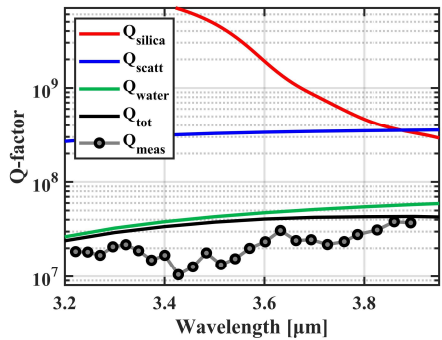

**Fig. 6 | Comparison of the measured and predicted Q-factor spectrum.** The total Q-factor ($Q_{tot}$) was predicted by adding its individually calculated loss components, namely $SiO_2$ absorption ($\propto 1/Q_{silica}$), roughness scattering ($\propto 1/Q_{scatt}$), and absorption from surface-adsorbed water ($\propto 1/Q_{water}$), and taking its inverse.

**Optical power measurement in the mid-IR SBS experiments.** Accurately measuring the optical power dynamics at mid-IR, especially for low threshold power on-chip devices, is difficult due to the lack of sensitive power meters (detectable power > ~1 µW) and low loss fiber components. To resolve this issue, we instead measured the waveguide in-coupled power using an ultra-sensitive photodetector (PD, VIGO Photonics PVI-4TE-5) placed at the waveguide transmission port (Figure 3d). The obtained PD voltage was converted to waveguide in-coupled power by the following steps. First, the PD responsivity was calibrated by comparing the PD voltage with a reference mid-IR power meter (Thorlabs S180C) readout. The PD voltage measured behind a pre-calibrated OD2 ND filter was plotted as a function of the optical power measured in front of the ND filter, giving a responsivity $1.73 \times 10^4$ V/W. Next, the net optical loss from the waveguide output facet to the photodetector, including the waveguide-to-fiber out-coupling loss and loss from free-space optics, was measured. The waveguide in-coupled power was calculated by dividing the PD transmission voltage by the calibrated responsivity and multiplying it with the measured loss factor. The output Stokes wave power was measured by multiplying the OSA power readout with the waveguide-to-OSA loss, obtained in the same manner.

During measurement, the waveguide-to-PD and waveguide-to-OSA losses constantly change due to the drift in waveguide-to-fiber out-coupling efficiency. This causes these two loss factors to deviate from the calibrated values on the same scale. Consequently, the measured input and Stokes power (x and y axes in Fig. 5c), obtained using these two factors, will also deviate from reality on the same scale. This causes the discrepancy between the measured and theoretical Stokes power dynamics in Figure 5c, represented by the scaling factor β = 1.43.

**Estimation of the Schawlow-Towns linewidth.** The Schawlow-Towns linewidth for the clamped Stokes wave was calculated using the following two equations[44,50]:

$$\Delta \nu_0 = \left(\frac{\Gamma}{\gamma + \Gamma}\right)^2 \frac{\hbar \omega_L^3 n_{th}}{4\pi Q_T Q_{Ex} P_{SBL}}, \qquad P_{SBL,clamp} = \frac{1}{g} \frac{\hbar \omega_L^3}{Q_T Q_{Ex}}$$

where $\hbar$ is the Plank constant, $\omega_L$ is the pump frequency, $n_{th} = 1693$ is the number of thermal phonon quanta calculated from the Bose-Einstein statistics, $Q_T (Q_{Ex})$ is the total (external) Q-factor of the Brillouin cavity, and $P_{SBL}$ is the power of the generated Stokes wave. The power of the clamped Stokes wave is inversely proportional to the Brillouin amplification rate per pump photon $g = \hbar \omega_L v_g^2 G_B / L = 1.3868$ Hz, where $v_g$ is the group velocity and $L$ is the physical cavity length[44]. Inserting the clamped power into the linewidth equation leads to the following simplified form:

$$\Delta \nu_{0,clamp} = \left(\frac{\Gamma}{\gamma + \Gamma}\right)^2 \frac{g}{4\pi} n_{th} = 83.45 \text{ Hz}$$

## Data availability

The data that support the findings of this study are available from the corresponding author upon reasonable request.

## Code availability

The data that support the findings of this study are available from the corresponding author upon reasonable request.

## References

51. Lee, H. *et al.* Chemically etched ultrahigh-Q wedge-resonator on a silicon chip. *Nature Photon* **6**, 369–373 (2012).
52. Choi, D.-Y. *et al.* Photo-induced and Thermal Annealing of Chalcogenide Films for Waveguide Fabrication. *Physics Procedia* **48**, 196–205 (2013).
53. Kim, D. *et al.* Two-point coupling method to independently control coupling efficiency at different wavelengths. *Opt. Lett., OL* **47**, 106–109 (2022).
54. Kita, D. M., Michon, J., Johnson, S. G. & Hu, J. Are slot and sub-wavelength grating waveguides better than strip waveguides for sensing? *Optica, OPTICA* **5**, 1046–1054 (2018).
55. Johnson, S. G. *et al.* Roughness losses and volume-current methods in photonic-crystal waveguides. *Appl. Phys. B* **81**, 283–293 (2005).

## Acknowledgements

This work was supported by the Samsung Research Funding & Incubation Center of Samsung Electronics under Project Number SRFC-IT1801-03. It was also supported by Institute for Information & communications Technology Planning&Evaluation grant (No. RS-2023-00223497) and the National Research Foundation of Korea grant (No. 2023R1A2C2004472), which are funded by the Korea government(MSIT). We thank Minsung Kim for taking microcavities image. H.L. acknowledge the support by KAIST Cross-Generation Collaborative Lab project.

## Author contributions

The study was conceived by K.K., D.S., D.C., and H.L. D.S. led the entire fabrication process. K.K., D.S., S.P., K.-H.K, D.K, and B.C performed the measurements. K.K., D.S., D.-H.K., B.S., and P.T.R analyzed the experimental results. R.W., Y.W., S.D., and X.W. provided the high purity chalcogenide glass ingots used for fabrication. All authors contributed to writing the paper. The project was supervised by D.C. and H.L.

## Competing interests

The authors declare no competing interests.